\documentclass[twocolumn,prc,floatfix,showpacs,preprintnumbers,nofootinbib,%
superscriptaddress]{revtex4}
\usepackage{mathrsfs}
\usepackage{amssymb}
\usepackage{amsmath}
\usepackage{graphicx}
\usepackage{hyperref}
\usepackage{mciteplus}

\newcommand{\xt}{{\mathbf{x}_T}}
\newcommand{\bt}{{\mathbf{b}_T}}

\newcommand{\ptt}{p_T} 
\newcommand{\ktt}{k_T} 

\newcommand{\ud}{\, \mathrm{d}}

\newcommand{\nc}{{N_\mathrm{c}}}

\newcommand{\cf}{C_\mathrm{F}}
\newcommand{\ca}{C_\mathrm{A}}

\newcommand{\nr}[1]{(\ref{#1})}

\newcommand{\ra}{R_A}
\newcommand{\rp}{R_p}

\newcommand{\tev}{\ \textrm{TeV}}
\newcommand{\gev}{\ \textrm{GeV}}
\newcommand{\fm}{\ \textrm{fm}}
\newcommand{\mb}{\ \textrm{mb}}

\newcommand{\qs}{Q_\mathrm{s}}
\newcommand{\qsprime}{Q_\mathrm{s}'}
\newcommand{\qsprimep}{Q_{\mathrm{s}p}'}
\newcommand{\qsprimea}{Q_\mathrm{sA}'}
\newcommand{\qsa}{Q_{\mathrm{s}A}}
\newcommand{\qsp}{Q_{\mathrm{s}p}}
\newcommand{\lqcd}{\Lambda_{\mathrm{QCD}}}
\newcommand{\as}{\alpha_{\mathrm{s}}}

\newcommand{\fig}{Fig.~}
\newcommand{\figs}{Figs.~}
\newcommand{\eq}{Eq.~}

\newcommand{\eqs}{Eqs.~}

\newcommand{\npart}{{N_\mathrm{part}}}
\newcommand{\nch}{{N_\mathrm{ch}}}

\newcommand{\xbj}{{x}}

\newcommand{\sigmap}{{ \sigma^\textrm{p}_\textrm{dip} }}

\newcommand{\dsigmap}{{\frac{\ud \sigma^\textrm{p}_\textrm{dip}}{\ud^2 \bt}}}
\newcommand{\dsigmaa}{{\frac{\ud \sigma^A_\textrm{dip}}{\ud^2 \bt}}}

\newcommand{\dsigma}{{\frac{\ud \sigma_\textrm{dip}}{\ud^2 \bt}}}

\begin{document}

\author{T. Lappi}
\affiliation{
Department of Physics, %
 P.O. Box 35, 40014 University of Jyv\"askyl\"a, Finland}
\affiliation{
Helsinki Institute of Physics, P.O. Box 64, 00014 University of Helsinki,
Finland}

\title{
Energy dependence of the saturation scale and 
the charged multiplicity in pp and AA collisions
}


\begin{abstract}
A natural framework to understand the energy dependence of bulk
observables from lower energy experiments to the LHC
is provided  by the Color Glass Condensate, which leads to 
a ``geometrical scaling'' in terms
of an energy dependent saturation scale $\qs$.
The  measured charged multiplicity, however, 
seems to grow faster ($\sim \sqrt{s}^{0.3}$) in nucleus-nucleus
collisions than it does for protons ($\sim \sqrt{s}^{0.2}$), violating
the expectation from geometric scaling. We argue that this difference
between pp and AA collisions
can be understood from the effect of DGLAP evolution on 
the value of the saturation scale, and is consistent with 
gluon saturation observations at HERA. 
\end{abstract}

\maketitle

\section{Introduction}

The unprecedented high energies of the LHC proton and nuclear beams
provide us with new experimentals tests of
QCD dynamics at high energy. On a fundamental level we know that also 
the bulk properties of the collision system such as the momentum
spectra and correlations of all produced hadrons must follow
from QCD. How this happens and to  
what extent the process can be understood  in a weak-coupling 
approximation is still an open issue. However, the phenomenological success of 
the Color Glass Condensate (CGC, for reviews
see e.g.~\cite{Iancu:2003xm,*Weigert:2005us,*Gelis:2010nm,*Lappi:2010ek})
makes one optimistic that this could in fact be possible.
In the CGC framework the small $x$ degrees of freedom that
dominate bulk particle production in hadronic collisions 
are described as nonperturbatively strong classical color
fields. This picture leads naturally to the concept
of gluon saturation and the \emph{saturation scale}
$\qs$ as the dominant transverse momentum scale determining
both the magnitude and the space and time dependence 
(i.e. the momenta of the gluons) of the small $x$ gluon fields.

The first LHC observable, in both proton proton and nucleus-nucleus
collisions, to give us information about  QCD dynamics at high energy
is the charged particle 
multiplicity~\cite{Aamodt:2010ft,*Aamodt:2010pp,Khachatryan:2010xs,Aamodt:2010pb}.
As we shall discuss in more detail in Sec.~\ref{sec:qsnch},
the charged multiplicity is, to a very good approximation, proportional
to $\qs^2$. Thus the energy dependence of the multiplicity is an experimental probe
of the $x$ dependence of the saturation scale $\qs^2$, separately
for nucleons and nuclei. The simplest 
model-independent way to see this is to realize that gluon saturation
turns particle production into a one scale problem, with 
$\qs$ as the only dimensionful scale apart from the size of the 
system. All bulk quantities, such as the multiplicity or the transverse energy,
and correlations in the system can be understood in this way up to  
normalization constants parametrically of order 1. 

In the CGC framework the energy dependence of the saturation scale
follows from the 
JIMWLK~\cite{Jalilian-Marian:1997xn,*Jalilian-Marian:1997jx,*Jalilian-Marian:1997gr,%
*Jalilian-Marian:1997dw,*JalilianMarian:1998cb,*Iancu:2000hn,%
*Iancu:2001md,*Ferreiro:2001qy,*Iancu:2001ad,*Mueller:2001uk} 
 renormalization group equation or 
its mean field, large $\nc$ approximation, the BK 
equation  \cite{Balitsky:1995ub,*Kovchegov:1999yj}.
At an intermediate scale, typically taken as $x\approx 0.01$,
where one would start the BK or JIMWLK evolution, the typical 
nuclear saturation
scale could be estimated as $\qsa^2 \approx c A^{1/3}\qsp^2$ with 
$c \sim A^{2/3}\rp^2/\ra^2$ a constant somewhat smaller than 1
(for a more detailed discussion see e.g.~\cite{Kowalski:2007rw}).
With a fixed coupling constant leading order JIMWLK or BK 
evolution would preserve the value of $\qsa/\qsp$
at all energies. Including
the running of $\as$, however, changes this picture.
At asymptotically high energies running coupling BK or JIMWLK evolution
leads to a saturation scale that is  independent of 
$A$~\cite{Mueller:2003bz} 
and therefore grows  more slowly for nuclei than for nucleons.

\begin{figure}[tb]
\includegraphics[width=0.5\textwidth]{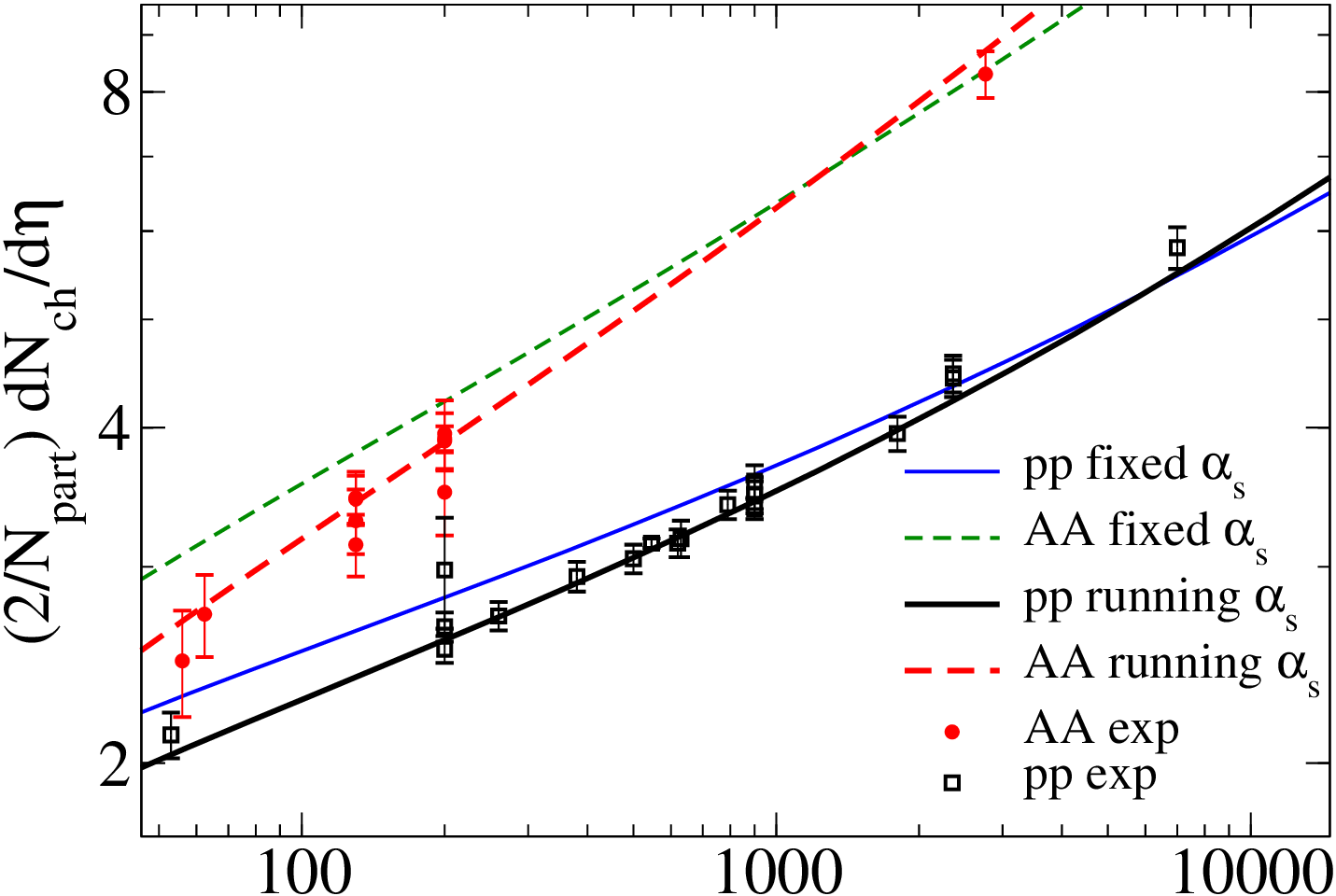}
\caption{
The charged multiplicity in pp and AA collisions estimated from 
the saturation scale in the IPsat model. Experimental datapoints from 
Refs.~\cite{Alner:1986xu,*Albajar:1989an,*Abe:1989td,Abelev:2008ez,Khachatryan:2010xs,Aamodt:2010ft,*Aamodt:2010pp} 
for pp and Refs.~\cite{Adler:2001yq,*Back:2000gw,*Bearden:2001xw,*Bearden:2001qq,*Back:2002uc,*Adler:2004zn,Abelev:2008ez,Aamodt:2010pb}
for AA collisions.
For details on relating $\qs$ to the charged multiplicity 
see discussion in Sec.~\ref{sec:qsnch}.
} \label{fig:ipsat}
\end{figure}

The arguments outlined above would lead to a primary gluon 
multiplicity that would have the same energy dependence
in pp and AA collisions (fixed coupling) or that grows more slowly
in AA than pp collisions (running coupling at asymptotical energies).
The trend seen in the data on the final charged multiplicity
is, however, the opposite one, posing a puzzle for attempts 
to understand it on the basis of gluon saturation.
We argue in this paper that at least a part of the explanation 
lies in transient effects that do not follow directly from
BK/JIMWLK evolution.
We will show that, due to effects of the nuclear 
geometry and of DGLAP evolution, the growth of 
$\qs$ with energy can actually be faster in nuclei than in 
protons. Thus a single parametrization of the
dipole cross section can be in good agreement with the basic features of the
experimental data in both protons and nuclei, as shown in \fig\ref{fig:ipsat}.
We will first discuss the energy dependence in 
different fits to HERA data in Sec.~\ref{sec:endep} before
looking more closely at what happens in two particular 
dipole cross section parametrizations, ``IPsat'' in Sec.~\ref{sec:ipsat}
and ``bCGC'' in  Sec.~\ref{sec:bcgc}.
We will then, in Sec.~\ref{sec:qsnch}, discuss in more detail the 
the relation between the initial gluonic and
final charged hadron multiplicity in nuclar 
and proton collisions.

\section{Energy dependence of the saturation scale}
\label{sec:endep}

Let us now discuss what is known about the value of the saturation scale
based on fits to HERA data
There exists by now a large amount of  different saturated
parametrizations of the 
 dipole cross section, mostly fit
to HERA or in some cases~\cite{Kharzeev:2004yx,*Dumitru:2005gt} 
to RHIC dAu data. Usually  the impact parameter dependence
is factorized into a Gaussian profile multiplying the
dimensionless scattering amplitude; this is the case
in the well known GBW~\cite{Golec-Biernat:1998js}
and IIM~\cite{Iancu:2003ge} parametrizations and the
more recent solutions of the running coupling BK 
equation~\cite{Albacete:2009fh}. In these cases the 
fit seems to favor an $x$-dependence of the saturation
scale that is faster than the observed energy dependence
of the charged multiplicity in pp collisions. This is the case
also for the BK evolution studied in~\cite{Albacete:2009fh}, where
the evolution speed is not a fit parameter, but follows 
directly from the evolution equation itself.
Thus the value $\lambda\approx 0.3$ obtained from 
a simple power law fit
$\ud \nch/\ud \eta \sim \sqrt{s}^{\lambda},$
 to the energy dependence of the 
multiplicity in AA collisions is in good agreement
with the evolution speed obtained e.g. in the 
GBW~\cite{Golec-Biernat:1998js} fit to HERA data. Although
some saturation calculations  slightly underpredicted
the AA multiplicity at the LHC, they still correctly
predict a stronger growth than in pp collisions.
The more recent application of running coupling BK evolution,
with a more detailed inclusion of the nuclear geometry,
reproduces the ALICE multiplicity data 
perfectly~\cite{Albacete:2010ad}.

On the other hand, it appears that in dipole cross section
parametrizations where the impact parameter dependence is not 
factorized out but included in the saturation scale itself, fits 
to HERA data prefer a slower increase of $\qs$ with $x$.
This is the case in both of the parametrizations, 
IPsat~\cite{Kowalski:2003hm,Kowalski:2006hc} and
bCGC~\cite{Kowalski:2006hc,Watt:2007nr} that we shall 
analyze in more detail in Secs~\ref{sec:ipsat}
and~\ref{sec:bcgc}. This is due to the functional
form that intertwines the $r$ and $b$ dependence of the 
dipole cross section in such a way that the proton is
allowed to grow with energy; leading to a growing
total DIS cross sections even with a slower increase
of the saturation scale. This growth is
consistent with the increase
of the total pp cross section with energy and the $t$-dependence
of diffractive vector meson production at HERA. 
As discussed e.g. in ~\cite{Kowalski:2008sa}, this is the
kind of parametrization that one would generally prefer on
theoretical grounds, since it causes the dipole cross 
section to saturate towards the correct black disk limit also
for $b\neq 0$. However, it is not obvious if the particular functional
form chosen in these parametrizations is the correct one.
The slower growth of $\qs$ in the IPsat and bCGC parametrizations
has, in $\ktt$-factorized calculations, yielded a good agreement
with the charged multiplicity in LHC proton-proton  
collisions~\cite{Levin:2010zy}. Also the multiplicity distributions
and $\ptt$-spectra in pp~collisions have recently been analyzed in the
$\ktt$-factorization approach~\cite{Tribedy:2010ab}.

In conclusion, among the different CGC fits to HERA data there
are ones that explain the multiplicity in pp collisions, and others
that give a good description of the multiplicity in AA collisions.
Neither the pp or AA multiplicity data separately is thus an indication
against gluon saturation, but the apparent failure to describe
both with the same parametrization is problematic. As we shall
now see, this is in fact not the case. The IPsat parametrization,
and also the bCGC (with an additional assumption that we will discuss), 
 can in fact describe
the $\sqrt{s}$ dependence of both the pp and AA multiplicities.

\begin{figure}[tb]
\includegraphics[width=0.5\textwidth]{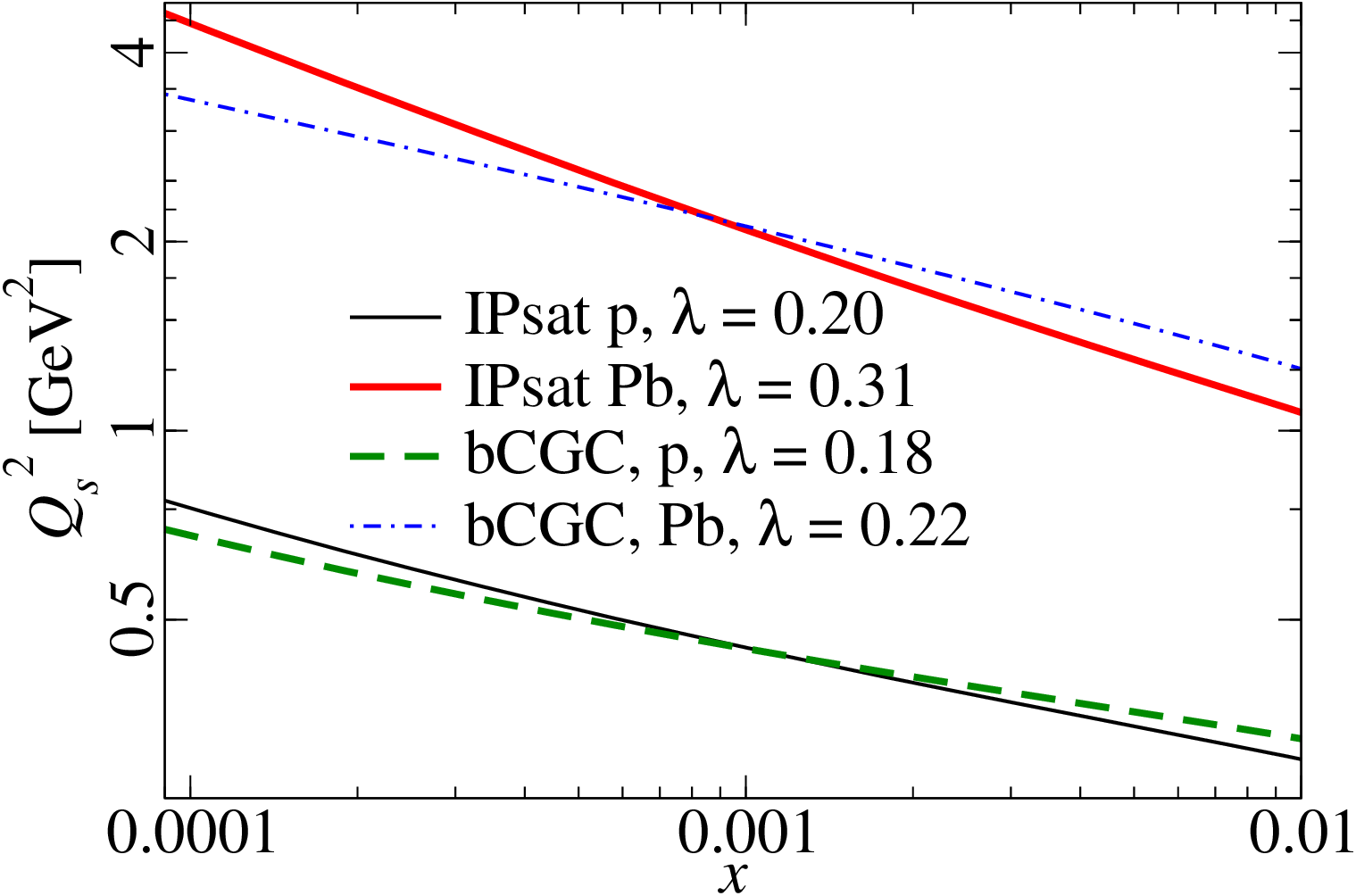}
\caption{
The saturation scale (adjoint representation) at the median impact parameter 
$b_\textrm{med}$ in a lead nucleus and a proton in the IPsat 
and bCGC models, as discussed in the text. Also shown are
the values $\lambda$ obtained by fitting an exponential
$\qs^2 = ax^{-\lambda}$ to the curves.
} \label{fig:qsvsx}
\end{figure}

\begin{figure}[tb]
\includegraphics[width=0.5\textwidth]{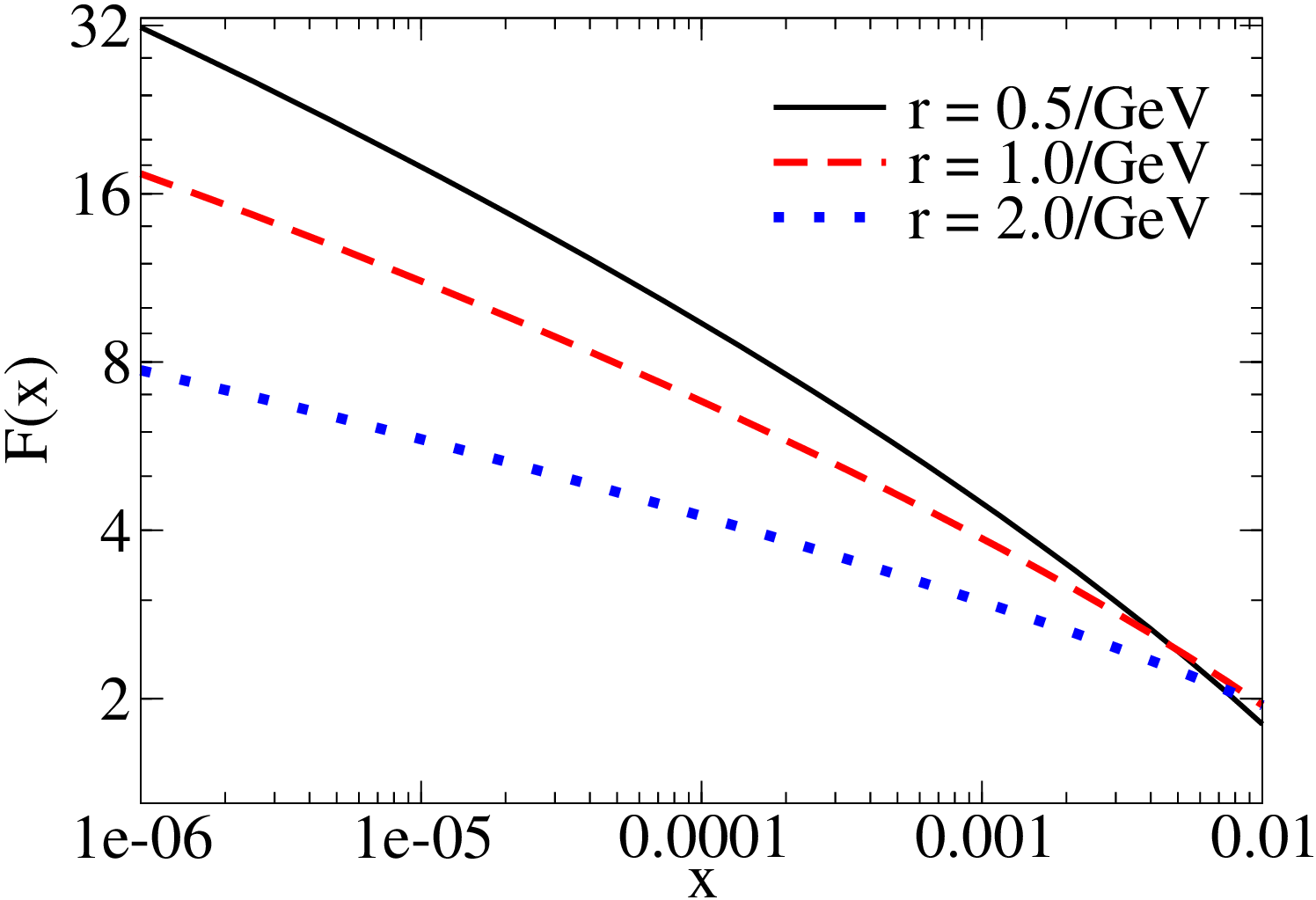}
\caption{
The function $F(x,r)$ of \eq\nr{eq:BEKW_F}, proportional to the DGLAP-evolved
gluon distribution, appearing in the IPsat model.
} \label{fig:gdist}
\end{figure}

\section{Nuclear effects in eikonalized DGLAP}
\label{sec:ipsat}

The IPsat model~\cite{Kowalski:2003hm,Kowalski:2006hc} 
is a modification of the idea
(see e.g.~\cite{Bartels:2002cj}) of including 
multiple scatterings and enforcing the black disk 
limit in the DIS cross section by exponentiating
(eikonalizing) a DGLAP-evolved gluon distribution. 
In the IPsat parametrization the impact parameter dependence is
included in the saturation scale (or DGLAP gluon distribution),
leading to the dipole cross section
\begin{equation}\label{eq:unfactbt}
\dsigmap(\bt,r,x)
 = 2\,\left[ 1 - \exp\left(- r^2  F(x,r) T_p(b)\right) 
\right].
\end{equation}
Here $T_p(b) = \exp\left(-b^2/2 B_p\right)/(2 \pi B)$ 
is the impact parameter profile function in the proton 
with $B_p=4.0\gev^2$ and $F$ is proportional to the 
gluon distribution
\begin{equation}
F(x,r) = 
\frac{ \pi^2 }{2 \nc} \as \left(\mu_0^2 + \frac{C}{r^2} \right) 
x g\left(x,\mu_0^2 + \frac{C}{r^2} \right)  .
\label{eq:BEKW_F}
\end{equation}
On a conceptual level, this formulation resums the multiple scatterings
off the small-$x$ gluons by assuming that they are independent and 
thus exponentiate. Strictly speaking it is not a ``CGC'' parametrization
in the sense that it does include the correlations
between the small-$x$ gluons in the target that are included
in the JIMWLK/BK equations. On the other hand it has the correct 
DGLAP behavior at large $Q^2$ (small $r$) and should be seen as a 
a good way to approach the saturation regime from the high $Q^2$ 
or large $x$ direction.

A generalization of \eq\nr{eq:unfactbt} to nuclei (including
fluctuations in the positions of the nucleons) is very 
straightforward: one replaces the thickness 
by a sum over $A$ nucleons as
\begin{equation}\label{eq:unfactbta}
\dsigmaa
 = 2\,\left[ 1 - \exp\left\{- r^2  F(x,r) \sum_{i=1}^{A}T_p(\bt-\bt_i)\right\} 
\right].
\end{equation}
This corresponds to treating the interactions with the separate
nucleons in the nucleus as independent; consistently with 
the eikonalization idea. In terms of the $S$-matrix of the 
dipole scattering off the target \eq\nr{eq:unfactbta} is 
equivalent to
\begin{equation}\label{eq:sfact}
S_A(r,\bt,x) = \prod_{i=1}^A S_p(r,\bt-\bt_i,x).
\end{equation}
The positions of the nucleons can then be averaged over 
to yield a Glauber-like averaged dipole cross section, written in 
the large $A$ approximation as
\begin{equation}
\dsigmaa
\approx 2\left[1-\exp\left\{-\frac{A T_A(\bt)}{2}\sigmap(r,x) \right\}\right],
\label{eq:nuc-dipole}
\end{equation}
where $\sigmap(r,x)$ is the nucleon dipole cross section of 
\eq\nr{eq:unfactbt} integrated over the impact parameter $\bt$.

The saturation scale $\qs$ characterizes the qualitative change
between the dilute color transparency region $r \to 0$ and
the black disk limit $\dsigma \to 2$ at large $r$.
We shall take here a model-independent definition of $\qs$ as the solution  of 
\begin{equation}\label{eq:defQs}
\dsigma(\xbj , r = 1/\qs(\xbj,\bt)) = 2(1-e^{-1/4}).
\end{equation}
The saturation scales defined as \eq\nr{eq:defQs} 
for a proton and a lead nucleus in the
IPsat parametrization are plotted in \fig\ref{fig:qsvsx}, multiplied
by the color factor $\ca/\cf$ appropriate for gluon production 
in pp or AA collisions.  We are
taking an effective value representing the average over the transverse plane
by taking the saturation scale at $b_\textrm{med}$, the median impact parameter
of the total DIS cross section (the value such that half the cross section
comes from $b<b_\mathrm{med}$). The difference between protons and nuclei
is striking: the energy dependence is $\qs^2 \sim x^{-0.31}$ for nuclei
and $\qs^2 \sim x^{-0.20}$ for protons, in a manner which immediately
evokes the $\sqrt{s}$-dependence of the charged multiplicity.

The reason for this difference lies in the behavior of the
DGLAP-evolved gluon distribution, whose $x$-dependence
gets steeper at higher $Q$ and, because $\qs$ grows with $A$, at
higher $A$. This feature was mentioned already in 
Ref.~\cite{Kowalski:2007rw}, although the discussion there is formulated
in terms of the $A$-dependence at fixed $x$ in stead of the $x$-dependence at
fixed $A$.
The function $F(x,r)$ of \eq\nr{eq:BEKW_F} is shown in \fig\ref{fig:gdist}.
The gluon distribution at the initial scale $\mu_0^2$ in the IPsat model
has a very mild $x$-dependence. The DGLAP evolution then drives the distribution
to become much steeper at higher scales which, because of the $A^{1/3}$ enhancement
of $\qs^2$ in nuclei, define the saturation region in a nucleus.

Let us try to estimate the saturation scales explicitly to illustrate
how this happens. From \eqs\nr{eq:unfactbt}, \nr{eq:BEKW_F} 
and \nr{eq:defQs} we can estimate the proton saturation scale 
at the center of the proton as
\begin{equation}\label{eq:qspimpl}
\qsp^2(b=0) = \frac{1}{4}\frac{F(x,r=1/\qsp(b=0))}{2\pi B_p}.
\end{equation}
Replacing, for purposes of illustration, the Woods-Saxon profile
by a theta function $T_A(b)\approx \theta(\ra-b)/(\pi \ra^2)$ we get for the 
saturation scale in a nucleus
\begin{equation}\label{eq:qsaest}
\frac{\sigmap(r=1/\qsa,x)/2}{\pi \ra^2}= \frac{1}{4}.
\end{equation}
Now we know that $\qsa>\qsp$ and therefore $r=1/\qsa$ 
is in the dilute region for
the proton dipole cross section. We can therefore take in 
\eq\nr{eq:qsaest} the small $r$ approximation 
 \begin{equation}
\sigmap(r=1/\qsa,x)/2 \approx \frac{1}{\qsa^2}F(x,r=1/\qsa)
\end{equation}
and thus
\begin{equation}\label{eq:qsaimpl}
\qsa^2(b=0) = \frac{1}{4}\frac{A F(x,r=1/\qsa(b=0))}{\pi \ra^2}.
\end{equation}
Equations \nr{eq:qspimpl} and \nr{eq:qsaimpl} are still implicit 
equations that must be solved to obtain the saturation scales. 
It is, however, easy to see that because $\qsa>\qsp$, the gluon 
distribution on the r.h.s. of \eqs\nr{eq:qspimpl} and~\nr{eq:qsaimpl}, 
and consequently the saturation scale on the
l.h.s., evolves more rapidly with energy.
We emphasize that this discussion is just an illustration of the origin 
of the different $x$-dependences in protons and nuclei, and the values in 
\fig\ref{fig:qsvsx} are obtained from the full expressions.

Finally, relating these saturation scales to the charged multiplicities
as discussed in Sec.~\ref{sec:qsnch} results in \fig\ref{fig:ipsat}. There are two curves
for protons and nuclei, differing by whether one keeps the coupling
$\as$ fixed at 0.33 or whether one allows it to run as $\as(\qs^2)$.
We see that the agreement with the experimental data is extremely good, 
considering the simplicity of this approach.

\section{Saturation scale for independent BK-evolved nucleons }
\label{sec:bcgc}

Whether the impact parameter profile is smooth (as in e.g.
\cite{Albacete:2009fh}) or fluctuating \cite{Albacete:2010ad}, 
running coupling BK evolution leads to a slower increase 
of $\qs$ for a higher initial value, i.e. for nuclei. 
Let us here consider an approximation that the 
individual nucleons evolve according to BK, and are then combined into a nucleus
using the assumption of independent scattering \eq\nr{eq:sfact}. Parametrically 
this ansatz could perhaps be justified at most in a moderate $x$ regime
where evolution does not yet happen coherently over the whole nucleus, 
and would certainly not be valid for asymptotically high energies. 
It is, however, interesting to see how a faster energy dependence 
in nuclei can arise also in this scenario.

To be more precise, the dipole cross section for a proton in the
bCGC parametrization~\cite{Kowalski:2006hc,Watt:2007nr} is:
\begin{align}
\dsigmap &= 2\,  {\cal N}_0 \left(\frac{r\qsprime}{2}\right)^{2\left(\gamma_s + 
{1\over \kappa \lambda Y}\ln\left(\frac{2}{r\qsprime}\right)\right)} &  \textrm{ for } & 
r\qsprime \leq 2\nonumber \\ 
&= 2 - 2\exp\left(-A\ln^2\left(Br\qsprime\right)\right) & \textrm{ for } & r\qsprime > 2 \, .
\label{eq:b-CGC}
\end{align}
The saturation scales $\qs$ and $\qsprime$ are conceptually the same quantity and 
their numerical values are of the same order, but we differentiate between them in order
to maintain our model independent definition of $\qs$ in \eq\nr{eq:defQs}.
The coefficients $A$ and $B$ in the can be determined uniquely 
from the condition that $\dsigmap$ and its first 
derivative with respect to $r\qsprime$ are continuous across $r\qsprime =2$. 
Here $Y = \ln(1/\xbj)$ is the rapidity, while $\gamma_s =0.628$ and $\kappa = 9.9$ 
(which quantifies the geometric scaling violations in \eq\nr{eq:b-CGC}) are 
obtained from leading logarithmic BFKL dynamics \cite{Iancu:2002tr}.
The impact parameter dependence of the proton saturation scale is introduced into the 
bCGC model in the form
\begin{equation}
\qsprime(\xbj,b) = \left(\frac{x_0}{\xbj}\right)^{\frac{\lambda}{2}}\left[\exp\left( 
- b^2/2 B_{\rm CGC}\right)\right]^{\frac{1}{2\gamma_s}} \gev \, .
\label{eq:qs-bCGC}
\end{equation}
The parameters $\lambda$, $x_0$, ${\cal N}_0$ and $B_{\rm CGC}$ are
 fit to the data, with the fit resulting in 
$\lambda=0.159$, $x_0=5.95\cdot10^{-4}$, ${\cal N}_0=0.417$ and 
$B_\mathrm{CGC}=5.5\gev^{-2}$.

We now use this parametrization for nuclei by assuming
that the scatterings off the nucleons are independent, i.e. assuming
\eq\nr{eq:sfact} which leads to the Glauber form for the avergare gluon
distribution~\nr{eq:unfactbta}. The resulting saturation scales,
at the median impact parameter $b_\mathrm{med}$ are plotted in 
\fig\ref{fig:qsvsx} for protons and nuclei. Due to the nontrivial functional
form of the parametrization, the evolution speed for protons turn out to 
be $0.18$, slightly larger than the parameter $\lambda$ in the 
parametrization, but still slower than in $b$-independent fits or in 
BK evolution. The nuclear saturation scale grows faster,
as $\qsa^2 \sim x^{-0.22}$. 

Let us try to understand this difference analytically in the similar
way as in Sec.~\ref{sec:ipsat}. It is easier
here to use the definition of the saturation scale $\qsprime$
that appears in the parametrization itself. Thus we have for the proton
\begin{equation}
\dsigmap(r=2/\qsprimep,x) =   2 \mathcal{N}_0
\end{equation}
and in the nucleus
\begin{multline}\label{eq:defqsprimea}
\dsigmaa\left(r=\frac{2}{\qsprimea},x\right) =  2 \mathcal{N}_0 \\
= 2\left[1-\exp\left\{-\frac{A T_A(b)}{2}\sigmap\left(r=\frac{2}{\qsprimea},x\right) 
\right\}\right].
\end{multline}
We again replace the Woods-Saxon 
distribution by a theta function $T_A(b)\approx \theta(\ra-b)/(\pi \ra^2)$.
Because at the nucleus saturation scale one is in the dilute
regime for the proton we can now approximate \eq\nr{eq:defqsprimea} by
\begin{equation}
\mathcal{N}_0 \approx
\frac{A T_A(b)}{2}\sigmap(r=1/\qsprimea,x).
\end{equation}
We now assume that the integral over the impact parameter
in the proton approximately factorizes into a constant 
$\sigma_0 \approx 2 \pi B_\mathrm{CGC}$ to get
\begin{equation}
\sigmap(r=1/\qsprimea,x) \approx 
2 \sigma_0 \mathcal{N}_0
 \left(\frac{\qsprimep}{\qsprimea}\right)^{2\left(\gamma_s + 
{1\over \kappa \lambda Y}\ln\left(\frac{\qsprimea}{\qsprimep}\right)\right)}, 
\end{equation}
leaving us with 
\begin{equation}\label{eq:delta}
\frac{A \sigma_0}{\pi \ra^2} = 
 \exp \left[ \left(\gamma_s + 
\frac{1}{ 2 \kappa \lambda Y } \Delta
\right) \Delta \right], 
\end{equation}
where we have denoted 
\begin{equation}
\frac{\qsprimea}{\qsprimep} \equiv e^{\Delta/2}.
\end{equation}
Since the l.h.s of \eq\nr{eq:delta} is independent of energy, it
is obvious that $\Delta$ must grow with the energy (i.e. with $Y$).
Differentiating with respect to $Y$ gives
\begin{equation}
\Delta'(Y) = \frac{\Delta^2}{2 \kappa \lambda Y^2(\gamma_s + 
\frac{\Delta}{\kappa \lambda Y})}.
\end{equation}
Assuming $\Delta \approx  \ln A^{1/3} \approx 1.8$
and $Y = \ln(1000) \approx 7$ we get
$\Delta' \approx 0.03$. In terms of the saturation scales
this means that
$\qsprimea^2 \sim \qsprimep x^{-0.03},$ which explains most of the 
effect seen in \fig\ref{fig:qsvsx}.
The interpretation of this result is in fact the same as
in the IPsat case. In the bCGC parametrization there is a
logarithmic term in the exponent that violates geometric 
scaling. At smaller $Y$, i.e. larger $x$, the effective
anomalous dimension $\gamma_\mathrm{eff} = 
\gamma_s + \ln(2/(r\qsprime))/(\kappa \lambda Y)$ is 
larger, i.e. closer to 1; thus the $Q^2$-dependence of the
integrated gluon distribution is close to a logarithm.
At large $Y$ or small $x$ one recovers the anomalous dimension
$\gamma_s$, which leads to a much faster increase of the
integrated gluon distribution with $Q^2$. This is precisely
the scenario that lead to a faster growth of $\qsa$ in the
case of the IPsat model. Another way to see this is to rewrite
\eq\nr{eq:delta} as
\begin{equation}
\left(\frac{\qsprimea}{\qsprimep}\right)^2 \approx 
\left( \frac{A \sigma_0}{\pi \ra^2} \right)^\frac{1}{\gamma_\mathrm{eff}}
\sim 
\left( A^{1/3} \right) ^\frac{1}{\gamma_\mathrm{eff}}.
\end{equation}
At smaller $x$ $1/\gamma_\mathrm{eff}$ is larger, thus the nuclear
enhancement of $\qs$ is larger.

\begin{figure}[tb]
\includegraphics[width=0.5\textwidth]{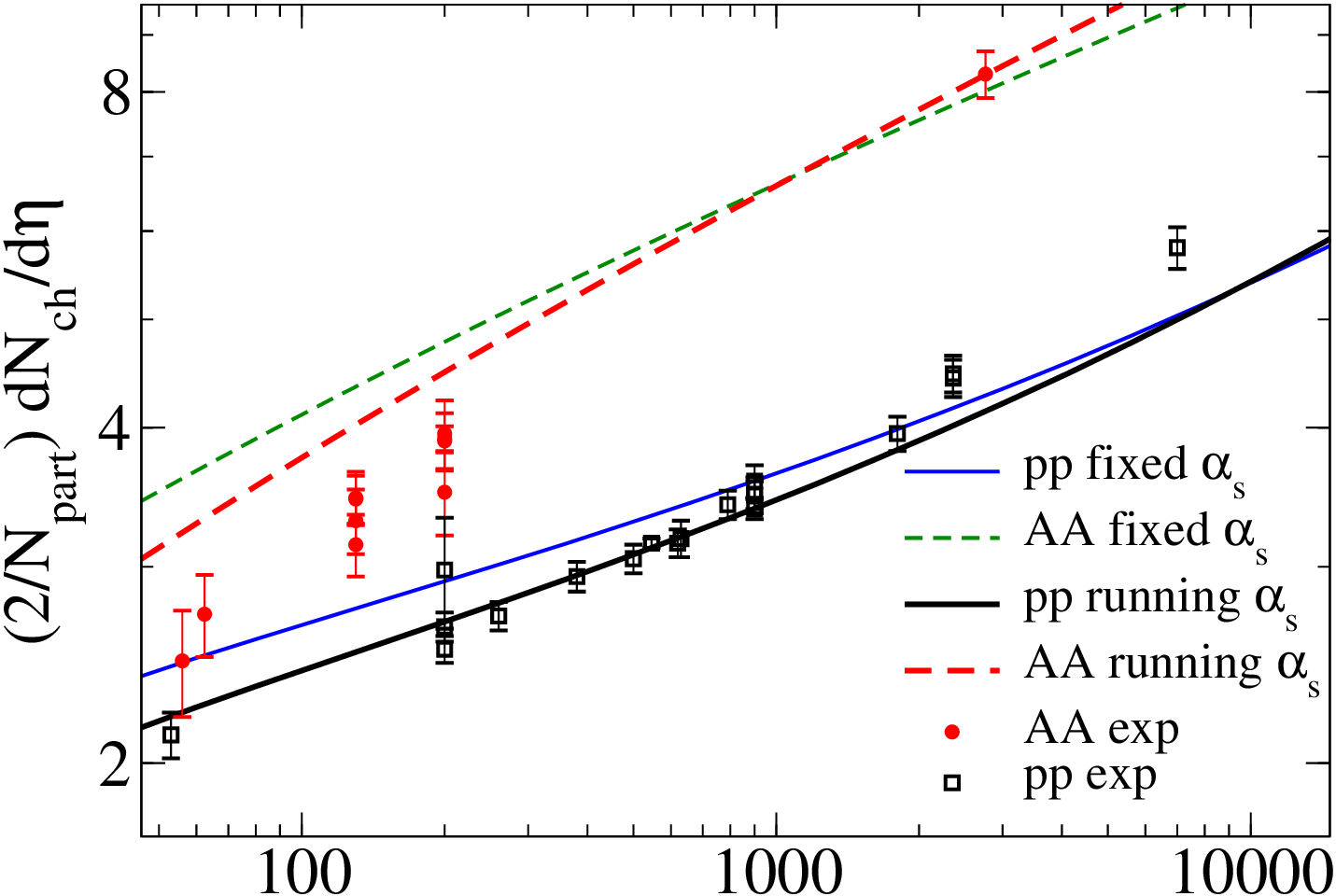}
\caption{
Energy dependence  of the charged multiplicity estimated using the 
bCGC saturation scale, extended to nuclei as described in
Sec.~\ref{sec:bcgc}
} \label{fig:bcgc}
\end{figure}

Again, using the  procedure described in 
Sec.~\ref{sec:qsnch} results in the estimates
for the charged multiplicity shown in \fig\ref{fig:bcgc}. 
While the agreement with experiental data is not as good as with the 
IPsat parametrization, the general trend of a faster increase in 
AA than in pp is still seen.

\section{Relation between $\qs$ and $N_\mathrm{ch}$}\label{sec:qsnch}

In any CGC calculation of gluon production in a collision of two hadronic
objects the initial gluon multiplicity depends on the saturation scale
parametrically in the same way. The theory of the CGC is based on weak coupling
calculations, and for the consistency of the framework one assumes
that $\as(\qs^2)\ll 1$. This means that $\qs$ is a semihard scale and we can 
assume that parametrically $\qs \gg \lqcd$. In the CGC, the saturation
scale also defines the correlation length of the system in the 
transverse plane, $\sim 1/\qs$. In the limit when a weak coupling 
CGC calculation is justified in the first place, the correlation length
is smaller than the size of the interaction region, 
$1/\qs^2 \ll \sigma$. Thus particle production happens \emph{locally}
in independent domains of size $\sim 1/\qs^2$ in the transverse plane.
This picture leads to a gluon multiplicity that can be written as a
local observable, where for dimensional reasons the number of 
gluons per unit area is proportional to the local $\qs^2$:
\begin{equation} \label{eq:libc}
\frac{\ud N_{\mathrm{init. }g}}{\ud^2 \xt \ud y} = c \frac{\cf \qs^2}{2 \pi^2 \as}.
\end{equation}
Here, following~\cite{Mueller:1999fp,*Mueller:2002kw,Kovchegov:2000hz}, we 
have introduced the ``gluon liberation coefficient'' $c$;
a nonperturbative dimensionless constant that is parametrically of
order $1$, but depends on the detailed spectrum of the produced gluons.
Its value in the MV model~\cite{McLerran:1994ni,*McLerran:1994ka,*McLerran:1994vd} has been determined using  Classical 
Yang-Mills simulations~\cite{Krasnitz:1998ns,*Krasnitz:2001qu,*Krasnitz:2003jw,Lappi:2003bi} 
to be $c\approx 1.1$ (see \cite{Lappi:2007ku} for a 
discussion of the CYM results parametrized 
in terms of  $c$ given here). 
As discussed e.g. in
Ref.~\cite{Lappi:2008eq} the value of $c$ remains very close to this value 
across a range of different models for 
the color charge distribution (or, equivalently, for the dipole cross section).
There is also an analytical calculation~\cite{Kovchegov:2000hz} 
of the liberation coefficient with the result
 $c \approx 2\ln2 \approx 1.4$. 

Phenomenological studies are often done using various 
$\ktt$-factorized approximations to compute the gluon spectrum. 
Although one can derive a $\ktt$-factorized formula for the dilute-dilute
``pp'' and dilute-dense ``pA'' cases, $\ktt$-factorization yields the
wrong gluon spectrum for the case of  dense-dense or ``AA'' 
scattering~\cite{Blaizot:2008yb,*Blaizot:2010kh,*Levin:2010zs}. Nevertheless, since
the only dimensionful scale in the problem is still $\qs$, also 
the result from $\ktt$-factorization 
can still be parametrized as \nr{eq:libc}, although
the value of the  coeffient $c$ is incorrect.

In proton proton collisions, the proportionality between the 
initial gluon and final charged hadron multiplicities is based on the
phenomenological success  of local parton-hadron 
duality (LPHD)~\cite{Dokshitzer:1991eq}. The working assumption
is that the final charged multiplicity
is proportional to the initial partonic one. The constant of proportionality 
is a property of the independent fragmentation of partons into hadrons and 
should thus be basically independent of collision energy or centrality.
In order to explain the slower growth of the charged multiplicity
in pp-collisions by a modification of the LPHD hypothesis one would
have to argue that as the energy increases, the number of charged
hadrons produced from a primary parton decreases, which does not
seem likely. Also the transverse area of the interaction region
should, if anything,  increase with energy. Therefore it seems that the
proton saturation scale should grow with energy at most as the observed
charged multiplicity, or even more slowly.

On the side of nucleus-nucleus collisions there is a wider range of 
plausible modifications to the relation between initial gluonic and observed
hadronic multiplicities. One possible starting point is the LPHD assumption as in 
pp collisions. The fragmentation could then be argued to lead 
to a larger ratio of the final charged hadron to the initial gluon 
multiplicity, e.g. due to the larger initial gluon momentum 
$\langle \ptt \rangle \sim \qs$; this is the argument in e.g.~\cite{Levin:2011hr}.
One potential problem for this approach comes from considering the first moment
of the particle spectrum, i.e. the total transverse energy. 
For the initial gluons the typical initial transverse momentum
is $\langle \ptt \rangle \sim \qs \sim \sqrt{(\ud N/\ud \eta)/S_\perp}$. 
A fragmentation process that leads to a larger final charged multiplicity
for AA than pp collisions should thus lead to 
$\langle \ptt \rangle /\sqrt{(\ud N/\ud \eta)/S_\perp}$ that 
decreases towards central collisions and towards higher energies.

 To be more explicit, let us
assume that in the LPHD picture one gluon produces $n$ charged 
particles after fragmentation. To account for the faster 
growth of the multiplicity with $\sqrt{s}$ in AA collisions, 
one would need $n_A > n_p$.
We now have for the initial gluons
\begin{eqnarray}
\langle \ptt \rangle_g & \sim&  \qs  \\
\frac{1}{S_\perp}\frac{\ud N_g}{\ud \eta } &\sim & \qs^2 
\end{eqnarray}
and for the final charged particles
\begin{eqnarray}
\langle \ptt \rangle_\mathrm{ch} & \sim& \qs /n  \\
\frac{1}{S_\perp}\frac{\ud \nch}{\ud \eta } &\sim & n \qs^2 ,
\end{eqnarray}
since transverse momentum must be conserved during fragmentation.
Now, if $n_A > n_p$, the scaled mean $\ptt$ 
\begin{equation} \label{eq:ptratio}
\frac{\langle \ptt \rangle_\mathrm{ch}}
{\sqrt{\frac{1}{S_\perp}\frac{\ud \nch}{\ud \eta }}} \sim \frac{1}{n\sqrt{n}}
\end{equation}
should be smaller in central AA collisions than for protons. This is indeed
seen in  the RHIC data~\cite{Abelev:2008ez}. However, if 
the increase in $n$ is due the larger  $\langle \ptt \rangle \sim \qs$
the ratio \nr{eq:ptratio} should also decrease with increasing collision energy.
Between $\sqrt{s}=62.4\gev$ and $\sqrt{s}=200\gev$ no such decrease is seen
at RHIC~\cite{Abelev:2008ez}, while no firm conclusions for higher energies
can yet be made from the LHC data.

The LPHD scenario is more or less based on neglecting all collective
effects even in AA collisions, and thus assuming that no quark gluon plasma is
formed. Ample experimental evidence  points to the contrary. 
In the extreme case of strong interactions among the produced gluons, they
thermalize into an isotropic plasma, which then expands in the transverse, 
and, more importantly, the longitudinal direction according to (nearly)
ideal hydrodynamical equations of motion. In the ideal hydrodynamical case
the entropy, and thus multiplicity, of the particles stays constant during
the evolution. This leads to the picture where the final (total) multiplicity
is, not only proportional, but nearly equal to the initial gluonic one.
During the hydrodynamical expansion of a locally isotropic 
system the mean transverse momentum (or energy per particle) 
decreases by a large amount due to $p\ud V$ work done pushing the expanding 
plasma down the beampipe. Radial flow developing during the evolution
also boosts the transverse momenta of the particles compared
to pp collisions, consistently with the observed increase with centrality
of $\langle \ptt \rangle$.
The drawback of this scenario is that, in spite of much work, we
do not have a quantitative theoretical understanding of the thermalization 
process.
This fast thermalization argument is, 
however, strangthened by the results of
explicit calculations of the initial gluon multiplicity in the CGC, which yield
an initial gluon multiplicity that is close to the final total multiplicity.
The initial transverse energy, on the other hand, is larger than the observed one
by a large factor, which would be consistent  with the hydrodynamical picture.

The true physical situation is most likely to lie somewhere between the
two extreme scenarios of LPHD and ideal hydrodynamics, 
with some entropy production and thus increase
in the multiplicity during the spacetime evolution of the plasma. 
However, based on this discussion it seems unlikely that final state effects
would solve the problem of a faster growth of the multiplicity with $\sqrt{s}$
in AA collisions than in pp.

Let us now use the assumption of
fast thermalization and ideal hydrodynamical expansion to obtain 
quantitative estimates for the charged multiplicity.
For central heavy ion collisions we take the following simple
estimate for the final charged multiplicity multiplicity: 
\begin{equation} \label{eq:aamulti}
\frac{2}{\npart }\frac{\ud N_\mathrm{ch}}{\ud \eta} 
\approx
\frac{2}{3} \frac{2}{\npart }\frac{\ud N_g}{\ud \eta} 
= 
\frac{2}{3} c \frac{\cf \qs^2(x)}{2 \pi^2 \as} \frac{2 S_\perp}{ \npart},
\end{equation}
Here the factor $2/3$ accounts for the fraction of charged particles 
of the total multiplicity  (to a first approximation $\pi^\pm,\pi^0$).
We take here $c=1.1$ as discussed above.
The typical transverse area $S_\perp$  per participant
pair is taken as the value estimated in central gold-gold collisions
by STAR~\cite{Abelev:2008ez} as
\begin{equation}
\frac{2 S_\perp}{\npart} \approx \frac{154 \fm^2}{0.5 \times 350}.
\end{equation}
The saturation scale as extracted from fits to DIS data 
depends on the momentum fraction $x$. In hadronic or heavy ion collisions
the corresponding variable is a ratio of  the transverse momentum
to the collision energy. 
The $x$ and $\qs(x)$ corresponding to each collision energy
$\qs(\sqrt{s})$  are solved from the relation 
\begin{equation}
x =  \frac{\qs(x)}{\sqrt{s}}.
\end{equation}

We emphasize that, unlike in typical $\ktt$-factorized calculations,
there is no arbitrary normalization factor to adjust here. Once the saturation 
scale is known, it determines both the normalization and the shape of the
$\ptt$ spectrum of the produced gluons. 
The mean transverse momentum  of the initial gluons is $\langle \ptt \rangle$
depends slightly more on the precise $\ktt$ dependence of the dipole 
cross section than the multiplicity.
For the case of the MV model the spectrum is relatively hard, with
$\langle \ptt \rangle \approx 1.3 \qs$ (following Refs.~\cite{Lappi:2003bi,Lappi:2007ku}).
For  RHIC this corresponds to $\langle \ptt \rangle \approx 1.5\gev$ 
for the intial gluons. With rapid thermalization and (nearly) ideal (nearly) boost 
invariant hydrodynamical evolution this is then reduced
(see e.g.~\cite{Eskola:1999fc}) by a factor $~3$ to
match the final observed transverse energy of around $0.5\gev$ per particle. At the 
$2.75 A \tev$ collision energy at the LHC
a similar estimate yields $\langle \ptt \rangle \approx 2.0\gev$ for the gluons in 
the initial state which would be similarly reduced by hydrodynamical evolution.

Now this can be contrasted with the estimate based on 
$\ktt$ factorization~\cite{Albacete:2010ad}, where, after adjusting the normalization constant 
to reproduce the RHIC multiplicity, one obtains at the LHC a transverse energy\footnote{
Note that this applies to the first version of Ref.~\cite{Albacete:2010ad}, 
and in a subsequent version there is a different 
normalization coefficient for the energy than for the multiplicity.}
$14\tev$ and a total multiplicity $\approx (3/2) 1600 =2400,$ i.e. a transverse energy of 
$6\gev$ per particle. Although a value for the transverse energy has not yet been released
by the LHC experiments, based on the spectra published in Ref.~\cite{Aamodt:2010jd}
it seems safe to say that this would require a reduction of at least a factor of 
6 between the initial gluons and the final state particles.
This failure to calculate both the initial multiplicity and transverse 
energy without adjusting both by an arbitrary normalization constant
follows from the incorrect gluon spectrum in $\ktt$-factorization.

For protons we simply replace \eq\nr{eq:aamulti} with an expression 
that leaves out the scaling by the number of participant pairs 
\begin{equation} \label{eq:ppmulti}
\frac{\ud N_\mathrm{ch}}{\ud \eta} = 
\frac{2}{3} c \frac{\cf \qs^2(x)}{2 \pi^2 \as} S_\perp.
\end{equation}
Here, adjusting the normalization to data as always with LPHD, 
we take as the transverse area by a constant $S_\perp=20\mb$
for fixed $\as$ and $S_\perp=24\mb$ for running $\as$. This includes
the conversion from gluons to final hadrons.
This summarizes the simple procedure used to arrive at the multiplicity
estimates in \figs\ref{fig:ipsat} and~\ref{fig:bcgc}.

\section{Conclusions}
\label{sec:conc}

The experimental data seems to 
be hinting that for bulk particle production at LHC energies one is
not yet far enough the asymptotic high energy, $\qs\gg \lqcd$,
regime for gluon saturation to work perfectly without additional 
finite $\sqrt{s}$ corrections. What these corrections are
remains still somewhat an open issue. One possibility is 
that soft confinement scale physics remains to be dominant
in pp collisions; in the weak coupling framework pursued in this
paper we have not been able to analyze this option.
 The other possibility, more encouraging in terms of prospects
for first principles understanding, is that the nuclear
saturation scale is large enough at the LHC to be sensitive
to the large $Q^2$ effects such as DGLAP evolution.
Note that also in the BK equation
the glowth of the saturation scale with energy is slower in 
the preasymptotic regime close to the initial condition. 
In phenomenological 
applications~\cite{Albacete:2007sm,Albacete:2009fh}
this preasymtotic slower growth is essential for agreement with 
experimental data. Thus the energy dependence at LHC energies is to a large
degree a consequence of the transverse momentum dependence in
the initial condition, not only of the evolution itself.

We have in this note shown that the different energy dependences
of the charged particle multiplicities in pp and AA collisions
can be understood in the framework of gluon saturation.
There is a well-tested and motivated impact parameter-dependent
parametrization, based on an eikonalized, DGLAP-evolved gluon 
distribution, that very accurately describes both 
pp and AA multiplicities. The difference between protons and nuclei comes,
in this parametrization, from effects of DGLAP evolution on the 
slope of the gluon distribution. These effects are not 
present in the pure JIMWLK/BK evolution formalism, whether
at fixed or running coupling. 
Fully understanding the physics at play here requires incorporating 
higher  transverse momentum physics not only into the initial condition,
but also  into nonlinear evolution itself.
Some steps this direction have already been 
taken~\cite{Triantafyllopoulos:2002nz,*Avsar:2009pv,*Avsar:2009pf}, 
but further investigation is needed to fully understand
the consequences for bulk particle production at the LHC.

\section*{Acknowledgements}
Discussions with R. Venugopalan are gratefully 
acknowledged.
This work  has been supported by the Academy of Finland, projects
126604 and 141555.

\newpage

\bibliography{spires}
\bibliographystyle{JHEP-2modM}

\end{document}